%% file: paper.tex
\documentclass[runningheads]{llncs}
\usepackage[T1]{fontenc}
\usepackage{graphicx}
\usepackage{longtable}
\usepackage{array,bbding}
\usepackage{cite}

\usepackage{paperstyle}

\def\orcidID#1{\href{http://orcid.org/#1}{\raisebox{-1.25pt}{\includegraphics{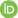}}}}

\begin{document}
\title{Synthesis Benchmarks for Automated Reasoning\thanks{This is the version of the paper accepted for CICM'25.}}
\titlerunning{Synthesis Benchmarks for Automated Reasoning}
\author{
M\'arton Hajdu\inst{1}\orcidID{0000-0002-8273-2613} \and
Petra Hozzov\'a\inst{2}\orcidID{0000-0003-0845-5811}  \and 
Laura Kov\'acs\inst{1}\orcidID{0000-0002-8299-2714} \and
Andrei Voronkov\inst{3,4} \and
Eva Maria Wagner\inst{1}\textsuperscript{(\Envelope)}\orcidID{0009-0006-3765-4130}
\and 
Richard Steven \v{Z}ilin\v{c}\'ik\inst{1}
}
\authorrunning{Hajdu, Hozzov\'a, Kov\'acs, Voronkov, Wagner, \v{Z}ilin\v{c}\'ik}
\institute{TU Wien, Vienna, Austria\\
eva.maria.wagner@tuwien.ac.at\and Czech Technical University, Prague, Czechia \and University of Manchester, Manchester, UK \and EasyChair, Manchester, UK} 
\maketitle              %

\renewcommand{\paragraph}[1]{\par\smallskip\noindent\textbf{#1}}

\begin{abstract}

Program synthesis is the task of constructing a program conforming to a given specification.
We focus on deductive synthesis, and in particular on synthesis problems with specifications given as $\forall\exists$-formulas, expressing the existence of an output corresponding to any input.
So far there has been no canonical benchmark set for deductive synthesis using the $\forall\exists$-format and supporting the so-called uncomputable symbol restriction. 
This work presents such a data set, composed by complementing existing benchmarks by new ones. Our data set is dynamically growing and should  motivate future  developments  in the theory and practice of automating synthesis. %

\end{abstract}

\section{Introduction} \label{section:introduction}
\input{sections/intro}

\section{Benchmark Format} \label{section:benchmark_format}
\input{sections/format}

\section{Benchmark Categories} \label{section:benchmark_categories}
\input{sections/categories}

\section{Evaluation and Experiments}\label{section:experinents}
To provide an insight into how challenging our benchmarks are, we ran {\sc cvc5}~\cite{cvc5,Reynolds19}, {\sc Synthesiz3}~\cite{Z3,synthesiz3}, and {\sc Vampire}~\cite{RecSynSat,SynSat,CAV13}. %
We used a time limit of 1 minute per benchmark, and ran the benchmarks on an Apple M1 Pro CPU with 32 GB memory.
We ran the solvers in the following configurations:
\begin{itemize}
\item {\sc cvc5}: version 1.2.1\footnote{\url{https://github.com/cvc5/cvc5/releases/}}, using the command {\tt cvc5 <benchmark.sl>}
\item {\sc Synthesiz3}: default configuration\footnote{\url{https://doi.org/10.5281/zenodo.16436706}}, using the command\\ {\tt python run\_synthesiz3.py <benchmark.smt2>}
\item {\sc Vampire}: {\tt synthesis-recursive} branch\footnote{\url{https://github.com/vprover/vampire/tree/synthesis-recursive}}, using the command\\ {\tt vampire -ind struct -qa synthesis -alasca off -tgt off -erd off -updr off -indc goal -nm 0 -ep off <benchmark.smt2>}\\
\end{itemize}

\medskip

The individual solvers solved the following numbers of benchmarks in each category:\footnote{The per-problem breakdown of the results is displayed in the appendix.}

\begin{center}
\tt
\begin{tabular}{p{1.8cm}>{\centering\arraybackslash}p{1.5cm}|>{\centering\arraybackslash}p{2cm}>{\centering\arraybackslash}p{2cm}>{\centering\arraybackslash}p{2cm}|>{\centering\arraybackslash}p{2cm}}
\hline
    {\rm Category} & {\rm Total} & {\sc cvc5} & {\sc Synthesiz3} & {\sc Vampire} & {\rm Total Solved}\\
    \hline
    NR\_LIA & 60 & 29 & 32 & 13 & 37\\
    NR\_LRA & 60 & 28 & 32 & 13 & 36\\
    NR\_NIA & 8 & 3 & 2 & 2 & 6\\
    NR\_NRA & 8 & 0 & 2 & 3 & 5\\
    NR\_UF & 20 & 4 & 4 & 20 & 20\\
    NR\_UFLIA & 12 & 5 & 10 & 3 & 10\\
    NR\_UFLRA & 12 & 7 & 8 & 2 & 9\\
    R\_UFDT & 110 & 3 & 0 & 17 & 17 \\ \hline
    altogether & 290 & 80 & 90 & 73 & 140 \\ \hline
\end{tabular}
\end{center}

\medskip

Given the expected growth of our benchmark set, as well as the agile development of state-of-the-art solvers, we emphasize that our experimental results  only depict the current capabilities of the respective synthesis tools.
We are confident that the solvers' performance will continue improving. 

\section{Conclusions} \label{section:conclusions}

We describe our program synthesis dataset where program specifications are given as  $\forall\exists$ first-order formulas.
Our dataset can be used for evaluating both the proving and synthesis abilities of automated reasoners in first-order logic, including reasoning with theories and induction.
Our dataset is planned to be further expanded by users and developers of reasoning tools, for %
example by extending our input format and considering benchmarks with multiple-invocation functions and  function composition benchmarks,
and with benchmarks using interpreted uncomputable symbols.

\subsubsection{Acknowledgements.}
This research was funded in whole or in part by the  ERC Consolidator Grant ARTIST 101002685, the ERC Proof of Concept Grant LEARN 101213411, the TU Wien Doctoral College SecInt, the FWF SpyCoDe Grant 10.55776/F85,  the WWTF grant ForSmart  10.47379/ICT22007, the Amazon Research Award 2023 QuAT, and by the European Union under the project ROBOPROX (reg. no. CZ.02.01.01/00/22\_008/0004590).

\subsubsection{Disclosure of Interests.}
The authors have no competing interests to declare that are relevant to the content of this article.

 \bibliographystyle{splncs04}
 \bibliography{bibliography}

\newpage
\appendix
\input{sections/appendix}

\end{document}

%% file: sections/intro.tex
Program synthesis constructs code  that is correct by design~\cite{ProgramSynthesis}, meaning that the synthesized code satisfies its given requirements. For the purpose of automating synthesis, such requirements are typically expressed in fragments of first-order logic (FOL) with theories, extending the use of automated reasoning engines towards deriving programs from the  correctness proofs of program requirements. 
This paper focuses on program synthesis using  specifications given as $\forall\exists$-formulas, expressing the existence of a program output corresponding to any input.\footnote{This specification class allows to specify single-invocation functions.}

\paragraph{Related work.} The automation of this instance of program synthesis has been recently  addressed by means of satisfiability modulo theory (SMT) solving~\cite{synthesiz3,Reynolds19} and/or first-order theorem proving~\cite{RecSynSat,SynSat,MannaWaldinger1980,MannaWaldinger1971}.
The approaches of~\cite{RecSynSat,SynSat,synthesiz3,Reynolds19} support different formats of program specifications with additional syntactic restrictions on the sought program: either (i) the syntax-guided synthesis (SyGuS) format~\cite{sygus-standard}, which supplies a context-free grammar used to construct the program, or (ii) specifications given as first-order logic $\forall\exists$-formulas, supplemented by a list of so-called uncomputable symbols that are forbidden to occur in the program~\cite{RecSynSat,SynSat}.
These format differences are significant enough that
SyGuS benchmarks cannot be expressed using the uncomputable framework of (ii), and further, dropping the context-free grammar often results in trivial benchmarks. 
On the other hand, while the format of (ii) allows to use specifications in full FOL combined with various theories, the SyGuS format does not support uninterpreted function/predicate symbols within FOL, limiting the expressivity of the SyGuS deductive synthesis benchmarks from the SyGuS competitions~\cite{sygus-comp}.
Overall, there is a lack of synthesis benchmarks requiring reasoning with theories, quantifiers and induction,
which would facilitate scaling deductive synthesis approaches.
This paper aims to bridge the synthesis formats of (i) and (ii), and 
introduces a dataset of synthesis problems that can be used both with uncomputable symbols and within extensions of SyGuS.

\paragraph{Our dataset.}
We collect 37 benchmarks from~\cite{RecSynSat,SynSat,synthesiz3}, and complement them with 253 new problems.
State of the art currently solves 140 benchmarks out of the 290 (see Section~\ref{section:experinents}). 
Our benchmarks are available online at:
\begin{center}
\url{https://github.com/vprover/vampire_benchmarks/tree/cicm2025/synthesis}
\end{center}%
Our dataset is continuously growing and invites  contributions from the communities of computer mathematics, automated reasoning, program verification, and others.

\medskip
\noindent{\it Input.} Problems in our dataset use program  specifications given  as a  pair of the specification formula and a set of uncomputable symbols $u_i$, namely 
\begin{equation}\label{eq:spec}
    \langle \underbrace{\forall x_1\in S_1,\dots, x_n\in S_n \exists y_1\in T_1,\dots,y_m\in T_m. \varphi[\bar{x}, \bar{y}]}_{\makebox[0pt]{\scriptsize specification formula}},\ 
    \underbrace{\{u_1,\dots,u_k\}}_{\makebox[0pt]{\scriptsize uncomputable symbols}}\rangle,
\end{equation}
where $\bar{x}$ denotes the set of \emph{input variables} $x_1,\dots,x_n$  of sorts $S_1,\dots,S_n$; $\bar{y}$ denotes the set of \emph{output variables}  $y_1,\dots,y_m$  of sorts $T_1,\dots,T_m$; $\varphi$ is  a many-sorted FOL formula, possibly using theories and recursive definitions, with the only free variables $\bar{x},\bar{y}$; and $u_1,\dots,u_k$ are \emph{uncomputable} symbols.
Intuitively, specification~\eqref{eq:spec} states the existence of $m$ functions with input variables $\bar{x}$, conforming to the specification formula, each independent of all of $u_1,\dots,u_k$.
If the specification formula $\varphi$ has the form $\psi\rightarrow\varphi'[\bar{x},\bar{y}]$, where $\psi$ is a closed formula, then $\varphi$  can  equivalently be written as
\begin{equation}\label{eq:spec-with-axioms}
    \psi \rightarrow \forall x_1\in S_1,\dots, x_n\in S_n \exists y_1\in T_1,\dots,y_m\in T_m. \varphi'[\bar{x}, \bar{y}],
\end{equation}
and in this case we call $\psi$ the \emph{background axioms and definitions}.

\medskip
\noindent{\it Output.} A solution for specification~\eqref{eq:spec} is
a tuple of terms $\langle r_1[\bar{x}], \dots, r_m[\bar{x}]\rangle$,
where
\(
\forall x_1\in S_1,\dots, x_n\in S_n. \varphi[x_1,\dots,x_n, r_1[\bar{x}],\dots,r_m[\bar{x}]]
\)
is valid in the underlying theory, and $r_i[\bar{x}]$ does not contain $u_j$ for all $i \in \{1,\dots,m\}$, $j \in \{1,\dots,k\}$.
We leave the output format up to individual synthesizers.
However, we note that the solution may use the $\itecons$ term constructor: for a formula $F$ and terms $s, t$ of the same sort the term $\ite{F}{s}{t}$ denotes the term $s$ if $F$ is true and $t$ otherwise.

\medskip
\noindent {\it Illustrative example.}
Let $\mathbb{N}$ denote the sort of natural numbers, and $\psi$ be the standard background definitions of $\leq, +, \cdot$ on $\mathbb{N}$.  Consider the problem from our dataset, 
\begin{equation}\label{eq:spec_ex}
    \langle \psi \rightarrow \forall x_1\in \N,\ x_2\in \N \ \exists y \in \N. \ x_1 + x_2 \leq y,\ 
    \{+\}\rangle,
\end{equation}
specifying a (possibly loose) upper bound on the sum of input variables $x_1,x_2$.
A possible solution for~\eqref{eq:spec_ex}, where 2 is an abbreviation for $s(s(0))$, is:
\begin{equation}\label{eq:sol_ex}
    \langle \ite{ x_1 \leq x_2}{x_2 \cdot 2}{x_1 \cdot 2}\rangle.
\end{equation}

%% file: sections/format.tex
We base our benchmark format on the SMT-LIB 2.7 syntax~\cite{SMTLIB2.7}.
As SMT-LIB is 
widely used by automated reasoners, our format makes automated reasoning naturally applicable to synthesis.
Our format, dubbed SMT-LIB-SYNU (SMT-LIB for synthesis with uncomputables), extends SMT-LIB with the following features:\medskip

\begin{enumerate}

\item Command {\tt assert-synth} defining
a synthesis specification formula of~\eqref{eq:spec} as %
\\
\phantom{\tt aaa}{\tt (assert-synth ((x1 S1)\dots(xn Sn)) ((y1 T1)\dots(yn Tm)) (}$\phi${\tt))}, \\
where $\mathtt{x1},\dots,\mathtt{xn}$ are the input variables, $\mathtt{y1},\dots,\mathtt{ym}$ are the output variables, and $\phi$ is the standard SMT-LIB encoding of the formula $\varphi$, with variables $x_i, y_j$ and sorts $S_i, T_j$ encoded as {\tt xi}, {\tt yj} and {\tt Si}, {\tt Tj}, respectively, for all $i\in\{1,\dots,n\}, j\in\{1,\dots,m\}$.

\item Command {\tt assert-claim}, asserting auxiliary lemmas, similarly to SMT-LIB's command {\tt assert}. The automated reasoner may decide to use these lemmas as axioms, or only use them after proving them as subgoals~\cite{LemmaDiscoveryAndStrategies} (see also Section~\ref{section:benchmark_categories}).

\item Option {\tt uncomputable} to annotate uncomputable symbols $u_1,\dots,u_n$ of~\eqref{eq:spec} as\\
\hspace*{6em} {\tt (set-option :uncomputable (u1 \dots{} un))}.\\
\end{enumerate}

\paragraph{Complementary format: SyGuS.} All of our benchmarks have been translated into the SyGuS 2.1 syntax~\cite{sygus-standard} using a Python script written by the authors.\footnote{Translation script {\tt synthesis/tools/smt2-to-sygus.py}  in our benchmark repository.}
We define computable recursive functions using the non-standard yet recommended command {\tt define-fun-rec}, already supported by some SyGuS solvers.
Since SyGuS does not allow declarations of uninterpreted functions, we encode functions that are declared but not defined, as well as uncomputable symbols, as follows.
We replace each declared function symbol and each uncomputable symbol by a (possibly higher-order) universally quantified variable of the same sort. SMT-LIB assertions over the symbols are encoded as SyGuS assumptions over the variables.
We add these variables as arguments to the signatures of the to-be-synthesized functions, and also pass the variables as function arguments in the constraint statement.
This allows the solver to use the variables corresponding to computable uninterpreted symbols in the solution. 

The only uncomputable symbols in our benchmarks are uninterpreted.
To express their uncomputability, it is sufficient to leave their corresponding higher-order variables out of the list of inputs for the function(s) to be synthesized.
We note that uncomputability of interpreted symbols can be expressed via SyGuS syntactic restriction -- by specifying a grammar allowing all computable symbols of the suitable sort.

%% file: sections/categories.tex
Our dataset currently contains 290 benchmarks. %
It consists of
(i) 180 \emph{non-recursive} problems and (ii) 110
\emph{recursive} problems over algebraic data types. We call benchmarks recursive if they contain recursively defined functions, as proving their specifications may require induction, leading to synthesizing recursive programs.

For selected benchmarks, we provide two versions, denoted by the file suffixes {\tt original} and {\tt assisted}.
The original benchmarks only contain the necessary axioms, while the assisted ones include auxiliary lemmas (using {\tt assert-claim}), which we found to be helpful for proving the existence of a solution.

\subsection{Non-Recursive Benchmarks}\label{section:benchmark_categories:nonrec}
Our non-recursive benchmarks are sub-categorized similarly to SMT-LIB logics, denoted with the prefix $\NR\texttt{\_}$.
For example, $\UF$ uses only uninterpreted functions and no theories, while $\LIA$ uses linear integer arithmetic.\footnote{see our benchmark repo for all our problems and their categorizations.}
The following $\UF$ example reasons  over  group theory. 
\begin{small}
\begin{examplebox}{Example of a \UF{} benchmark (\texttt{group-3-square\_commutativity\_assisted.smt2})}{4.3cm}
\begin{verbatim}
(set-logic UF)\end{verbatim}
  \exvspace
\begin{verbatim}
(declare-sort s 0)
(declare-fun inv (s) s)
(declare-fun op (s s) s)
(declare-const e s)\end{verbatim}
  \exvspace
\begin{verbatim}
; Group axioms: left inverse, left identity, associativity
(assert (forall ((x s)) (= (op (inv x) x) e)))
(assert (forall ((x s)) (= (op e x) x)))
(assert (forall ((x s) (y s) (z s))
    (= (op x (op y z)) (op (op x y) z))))
; Additional axiom to aid the solvers: right identity
(assert (forall ((x s)) (= (op x e) x)))\end{verbatim}
  \exvspace
\begin{verbatim}
; Synthesis task
(assert-synth ((x s) (y s)) ((z s))
    (=> (distinct (op x y) (op y x)) (distinct (op z z) e)))
\end{verbatim}
\tcblower
The problem uses a group with binary operation {\tt op}, inverse element function {\tt inv}, and neutral element {\tt e}.
It specifies a function with two inputs {\tt x} and {\tt y}, such that when {\tt op} is not commutative on {\tt x}, {\tt y}, then the square of the output is not {\tt e}. In mathematical notation this is $\forall x, y \ \exists z. \ x * y \neq y * x \rightarrow z * z \neq e$.

\end{examplebox}
\end{small}

\subsection{Recursive Benchmarks}\label{section:benchmark_categories:rec}
Our recursive benchmarks, denoted $\DT$, use the inductive datatypes $\nat$ of natural numbers;  %
$\lists$ of lists; %
and $\trees$ of binary trees. %
In some benchmarks, $\lists$ and $\trees$ use $\nat$ as the element type; in others, they use arbitrary types.
The following recursive benchmark  encodes example~$\eqref{eq:spec_ex}$ of Section~\ref{section:introduction}, using the $\nat$ datatype:
\begin{small}
\begin{examplebox}{Example of a \DT{} benchmark (\texttt{nat-upper\_bound\_of\_sum.smt2})}{4.3cm}
\begin{verbatim}
(set-logic UFDT)
\end{verbatim}
  \exvspace
\begin{verbatim}
; Datatype declarations
(declare-datatype nat ((zero) (s (s0 nat))))
\end{verbatim}
  \exvspace
\begin{verbatim}
; Function definitions
(define-fun-rec add ((x nat) (y nat)) nat
    (match y ((zero x) ((s y0) (s (add x y0))))))
(define-fun-rec mult ((x nat) (y nat)) nat
    (match y ((zero zero) ((s y0) (add (mult x y0) x)))))
(define-fun leq ((x nat) (y nat)) Bool (exists ((k nat))(= (add x k) y)))
\end{verbatim}
  \exvspace
\begin{verbatim}
; Synthesis task
(assert-synth ((x1 nat) (x2 nat)) ((y nat)) (leq (add x1 x2) y))
\end{verbatim}
\exvspace
\begin{verbatim}
(set-option :uncomputable (add))
\end{verbatim}
\tcblower
The inductive datatype $\nat$ has constructors $\tt{zero}$ and $\tt{s}$. The recursive functions $\tt{add}$ and $\tt{mult}$ are defined over $\nat$. The function $\tt{leq}$ is defined over $\nat$ using $\tt{add}$. The specification asks for an upper bound on the sum of two inputs wrt $\tt{leq}$.
\end{examplebox}
\end{small}
We give an additional example input file from our $\DT$ category that reasons over mixed datatypes: a $\lists$ containing $\nat$ elements.%
\vspace{-0.5em}
\begin{small}
\begin{examplebox}{Example of a \DT{} benchmark (\texttt{list-max\_elem.smt2})}{4.3cm}
\begin{verbatim}
(set-logic UFDT)
\end{verbatim}
  \exvspace
\begin{verbatim}
; Datatype declarations
(declare-datatype nat ((s (s0 nat)) (zero)))
(declare-datatype lst ((nil) (cons (cons0 nat) (cons1 lst))))
\end{verbatim}
  \exvspace
\begin{verbatim}
; Function definitions
(define-fun-rec add ((x nat) (y nat)) nat
    (match y ((zero x)
             ((s y0) (s (add x y0))))))
(define-fun leq ((x nat) (y nat)) Bool (exists ((k nat))(= (add x k) y)))
(define-fun-rec inlst ((x nat) (y lst)) Bool
    (match y ((nil false)
             ((cons y0 y1) (or (= x y0) (inlst x y1))))))
\end{verbatim}
  \exvspace
\begin{verbatim}
; Synthesis task
(assert-synth ((l lst)) ((n nat))
    (and (inlst n l) (forall ((k nat)) (=> (inlst k l) (leq k n)))))
\end{verbatim}
\tcblower
The datatype $\lists$ is defined containing elements of type $\nat$ with constructors $\tt{nil}$ and $\tt{cons}$. The recursive function $\tt{inlst}$ is defined over lists. The specification asks for a program that takes a list as input and returns the biggest element.
\end{examplebox}
\end{small}

%% file: sections/appendix.tex
\newcolumntype{R}[2]{%
    >{\adjustbox{angle=#1,lap=\width-(#2)}\bgroup}%
    l%
    <{\egroup}%
}
\newcommand*\rot{\multicolumn{1}{R{45}{1em}}}

\section{Appendix}
Here we show the per-benchmark results for {\sc cvc5}, {\sc Synthesiz3}, and {\sc Vampire}:\\[-2em]

\begin{center}
\tt
\small
\begin{longtable}{|l|>{\centering\arraybackslash}p{0.8cm}|>{\centering\arraybackslash}p{0.8cm}|>{\centering\arraybackslash}p{0.8cm}|}
    \multicolumn{1}{l}{\rm Benchmark} & \rot{\sc cvc5} & \rot{\sc Synthesiz3} & \rot{\sc Vampire} \\
    \hline \endhead
    \multicolumn{4}{c}{}\\[-0.95em] \hline
    \multicolumn{4}{|c|}{NR\_LIA}\\
    \hline\hline
    lower\_bound2 & \checkmark & \checkmark & \checkmark \\
    lower\_bound5 & \checkmark & \checkmark & \checkmark \\
    lower\_bound10 & \checkmark & \checkmark & - \\
    lower\_bound15 & \checkmark & \checkmark & - \\
    lower\_bound20 & \checkmark & - & - \\
    lower\_bound25 & - & - & - \\
    lower\_bound30 & - & - & - \\
    lower\_bound50 & - & - & - \\
    lower\_bound100 & - & - & - \\
    lower\_bound150 & - & - & - \\
    lower\_strict2 & \checkmark & \checkmark & \checkmark \\
    lower\_strict5 & \checkmark & \checkmark & - \\
    lower\_strict10 & \checkmark & \checkmark & - \\
    lower\_strict15 & \checkmark & - & - \\
    lower\_strict20 & \checkmark & - & - \\
    lower\_strict25 & - & - & - \\
    lower\_strict30 & - & - & - \\
    lower\_strict50 & - & - & - \\
    lower\_strict100 & - & - & - \\
    lower\_strict150 & - & - & - \\
    max2 & \checkmark & \checkmark & \checkmark \\
    max5 & \checkmark & \checkmark & \checkmark \\
    max10 & \checkmark & \checkmark & \checkmark \\
    max15 & \checkmark & \checkmark & \checkmark \\
    max20 & \checkmark & \checkmark & - \\
    max25 & - & \checkmark & - \\
    max30 & - & \checkmark & - \\
    max50 & - & \checkmark & - \\
    max100 & - & \checkmark & - \\
    max150 & - & - & - \\
    min2 & \checkmark & \checkmark & \checkmark \\
    min5 & \checkmark & \checkmark & \checkmark \\
    min10 & \checkmark & \checkmark & \checkmark \\
    min15 & \checkmark & \checkmark & \checkmark \\
    min20 & \checkmark & \checkmark & - \\
    min25 & - & \checkmark & - \\
    min30 & - & \checkmark & - \\
    min50 & - & \checkmark & - \\
    min100 & - & \checkmark & - \\
    min150 & - & - & - \\
    upper\_bound2 & \checkmark & \checkmark & \checkmark \\
    upper\_bound5 & \checkmark & \checkmark & - \\
    upper\_bound10 & \checkmark & \checkmark & - \\
    upper\_bound15 & \checkmark & \checkmark & - \\
    upper\_bound20 & - & - & - \\
    upper\_bound25 & - & - & - \\
    upper\_bound30 & - & - & - \\
    upper\_bound50 & - & - & - \\
    upper\_bound100 & - & - & - \\
    upper\_bound150 & - & - & - \\
    upper\_strict2 & \checkmark & \checkmark & \checkmark \\
    upper\_strict5 & \checkmark & \checkmark & - \\
    upper\_strict10 & \checkmark & \checkmark & - \\
    upper\_strict15 & \checkmark & - & - \\
    upper\_strict20 & \checkmark & - & - \\
    upper\_strict25 & - & - & - \\
    upper\_strict30 & - & - & - \\
    upper\_strict50 & - & - & - \\
    upper\_strict100 & - & - & - \\
    upper\_strict150 & - & - & - \\
    \hline\hline
    \multicolumn{4}{|c|}{NR\_LRA}\\
    \hline\hline
    lower\_bound2 & \checkmark & \checkmark & \checkmark \\
    lower\_bound5 & \checkmark & \checkmark & - \\
    lower\_bound10 & \checkmark & \checkmark & - \\
    lower\_bound15 & \checkmark & \checkmark & - \\
    lower\_bound20 & \checkmark & - & - \\
    lower\_bound25 & - & - & - \\
    lower\_bound30 & - & - & - \\
    lower\_bound50 & - & - & - \\
    lower\_bound100 & - & - & - \\
    lower\_bound150 & - & - & - \\
    lower\_strict2 & \checkmark & \checkmark & \checkmark \\
    lower\_strict5 & \checkmark & \checkmark & - \\
    lower\_strict10 & \checkmark & \checkmark & - \\
    lower\_strict15 & \checkmark & - & - \\
    lower\_strict20 & - & - & - \\
    lower\_strict25 & - & - & - \\
    lower\_strict30 & - & - & - \\
    lower\_strict50 & - & - & - \\
    lower\_strict100 & - & - & - \\
    lower\_strict150 & - & - & - \\
    max2 & \checkmark & \checkmark & \checkmark \\
    max5 & \checkmark & \checkmark & \checkmark \\
    max10 & \checkmark & \checkmark & \checkmark \\
    max15 & \checkmark & \checkmark & \checkmark \\
    max20 & \checkmark & \checkmark & - \\
    max25 & - & \checkmark & - \\
    max30 & - & \checkmark & - \\
    max50 & - & \checkmark & - \\
    max100 & - & \checkmark & - \\
    max150 & - & - & - \\
    min2 & \checkmark & \checkmark & \checkmark \\
    min5 & \checkmark & \checkmark & \checkmark \\
    min10 & \checkmark & \checkmark & \checkmark \\
    min15 & \checkmark & \checkmark & \checkmark \\
    min20 & \checkmark & \checkmark & - \\
    min25 & - & \checkmark & - \\
    min30 & - & \checkmark & - \\
    min50 & - & \checkmark & - \\
    min100 & - & \checkmark & - \\
    min150 & - & - & - \\
    upper\_bound2 & \checkmark & \checkmark & \checkmark \\
    upper\_bound5 & \checkmark & \checkmark & \checkmark \\
    upper\_bound10 & \checkmark & \checkmark & - \\
    upper\_bound15 & \checkmark & \checkmark & - \\
    upper\_bound20 & \checkmark & - & - \\
    upper\_bound25 & - & - & - \\
    upper\_bound30 & - & - & - \\
    upper\_bound50 & - & - & - \\
    upper\_bound100 & - & - & - \\
    upper\_bound150 & - & - & - \\
    upper\_strict2 & \checkmark & \checkmark & \checkmark \\
    upper\_strict5 & \checkmark & \checkmark & - \\
    upper\_strict10 & \checkmark & \checkmark & - \\
    upper\_strict15 & \checkmark & - & - \\
    upper\_strict20 & - & - & - \\
    upper\_strict25 & - & - & - \\
    upper\_strict30 & - & - & - \\
    upper\_strict50 & - & - & - \\
    upper\_strict100 & - & - & - \\
    upper\_strict150 & - & - & - \\
    \hline\hline
    \multicolumn{4}{|c|}{NR\_NIA}\\
    \hline\hline
    abs & \checkmark & - & - \\
    polynomial1 & - & - & \checkmark \\
    polynomial1\_3ord & - & - & - \\
    polynomial1\_3vars & - & - & - \\
    polynomial2 & - & - & \checkmark \\
    quotient-1 & - & \checkmark & - \\
    same-quotient & \checkmark & \checkmark & - \\
    signum & \checkmark & - & - \\
    \hline\hline
    \multicolumn{4}{|c|}{NR\_NRA}\\
    \hline\hline
    abs & - & - & - \\
    polynomial1 & - & - & \checkmark \\
    polynomial1\_3ord & - & - & - \\
    polynomial1\_3vars & - & - & - \\
    polynomial2 & - & - & \checkmark \\
    quotient-1 & - & - & \checkmark \\
    same-quotient & - & \checkmark & - \\
    signum & - & \checkmark & - \\
    \hline\hline
    \multicolumn{4}{|c|}{NR\_UF}\\
    \hline\hline
    UF/fu\_is\_a & \checkmark & \checkmark & \checkmark \\
    UF/u\_is\_a & \checkmark & \checkmark & \checkmark \\
    field/1 & - & - & \checkmark \\
    field/2 & - & - & \checkmark \\
    group/ex1-right\_inverse & - & - & \checkmark \\
    group/ex1-right\_inverse\_2 & - & - & \checkmark \\
     group/ex1-right\_inverse\_5 & - & - & \checkmark \\
    group/ex1-right\_inverse\_10 & - & - & \checkmark \\
    group/ex1-right\_inverse\_15 & - & - & \checkmark \\
    group/ex1-right\_inverse\_25 & - & - & \checkmark \\
    group/ex2-inverse\_of\_ixopiy & - & - & \checkmark \\
     group/ex2-inverse\_of\_ixopiy\_3 & - & - & \checkmark \\
    group/ex2-inverse\_of\_ixopiy\_5 & - & - & \checkmark \\
    group/ex2-inverse\_of\_ixopiy\_10 & - & - & \checkmark \\
    group/ex2-inverse\_of\_ixopiy\_15 & - & - & \checkmark \\
    group/ex2-inverse\_of\_ixopiy\_25 & - & - & \checkmark \\
    group/ex3-square\_commutativity\_assisted & - & - & \checkmark \\
    group/ex3-square\_commutativity\_original & - & - & \checkmark \\
    workshop/1 & \checkmark & \checkmark & \checkmark \\
    workshop/2 & \checkmark & \checkmark & \checkmark \\
    \hline\hline
    \multicolumn{4}{|c|}{NR\_UFLIA}\\
    \hline\hline
    between1 & \checkmark & \checkmark & \checkmark \\
    between2 & \checkmark & \checkmark & \checkmark \\
    between\_strict1 & \checkmark & \checkmark & \checkmark \\
    between\_strict2 & \checkmark & \checkmark & - \\
    equation1 & - & \checkmark & - \\
    equation2 & - & \checkmark & - \\
    equation3 & - & \checkmark & - \\
    knapsack1 & \checkmark & \checkmark & - \\
    knapsack2 & - & \checkmark & - \\
    knapsack3 & - & \checkmark & - \\
    knapsack4 & - & - & - \\
    knapsack5 & - & - & - \\
    \hline\hline
    \multicolumn{4}{|c|}{NR\_UFLRA}\\
    \hline\hline
    between1 & \checkmark & \checkmark & \checkmark \\
    between2 & \checkmark & \checkmark & \checkmark \\
    between\_strict1 & \checkmark & - & - \\
    between\_strict2 & - & - & - \\
    equation1 & \checkmark & \checkmark & - \\
    equation2 & \checkmark & \checkmark & - \\
    equation3 & \checkmark & \checkmark & - \\
    knapsack1 & \checkmark & \checkmark & - \\
    knapsack2 & - & \checkmark & - \\
    knapsack3 & - & \checkmark & - \\
    knapsack4 & - & - & - \\
    knapsack5 & - & - & - \\
    \hline\hline
    \multicolumn{4}{|c|}{R\_UFDT}\\
    \hline\hline
     tree/max-min-bound/max\_elem & - & - & - \\
    tree/max-min-bound/min\_elem & - & - & - \\
     tree/max-min-bound/upper\_bound & - & - & - \\
    tree/max-min-bound/lower\_bound & - & - & - \\
    tree/mirrored/1 & - & - & - \\
    tree/mirrored/2 & - & - & - \\
    tree/mirrored/3 & - & - & - \\
    tree/mirrored/4 & - & - & - \\
    list/assoc/left\_4\_vars & - & - & - \\
    list/assoc/left\_5\_vars & - & - & - \\
    list/assoc/left\_minimal & - & - & \checkmark \\
    list/assoc/right\_4\_vars & - & - & \checkmark \\
    list/assoc/right\_5\_vars & - & - & \checkmark \\
    list/assoc/right\_minimal & - & - & \checkmark \\
    list/len/concat\_minimal & - & - & \checkmark \\
    list/len/concat\_3\_vars & - & - & - \\
    list/max-min-bound/max\_elem & - & - & - \\
    list/max-min-bound/min\_elem & - & - & - \\
    list/max-min-bound/upper\_bound & - & - & - \\
     list/max-min-bound/lower\_bound & - & - & - \\
    list/prefix-suffix/prefix-given-suffix\_assisted & - & - & - \\
    list/prefix-suffix/prefix-given-suffix\_original & - & - & - \\
    list/prefix-suffix/suffix-given-prefix\_assisted & - & - & - \\
    list/prefix-suffix/suffix-given-prefix\_original & - & - & - \\
    list/prime\_fac & - & - & - \\
    list/prod\_of\_lists & - & - & - \\
    list/projection/first-elem\_assisted & - & - & \checkmark \\
    list/projection/first-elem\_original & \checkmark & - & \checkmark \\
    list/projection/kth\_elem & - & - & - \\
    list/projection/last\_elem & - & - & - \\
    list/rev/1\_assisted & - & - & - \\
    list/rev/1\_original & - & - & - \\
    list/rev/2\_assisted & - & - & - \\
    list/rev/2\_original & - & - & - \\
    list/rev/3 & - & - & - \\
    list/rev/4 & - & - & - \\
    list/rev/5 & - & - & - \\
    list/rev/6 & - & - & - \\
    list/rev/7 & - & - & - \\
    list/sort/1 & - & - & - \\
    list/sort/2 & - & - & - \\
    list/sort/3 & - & - & - \\
    list/sort/4 & - & - & - \\
    list/sort/5 & - & - & - \\
    list/sum/1\_assisted & - & - & - \\
    list/sum/1\_original & - & - & - \\
    list/sum/2\_assisted & - & - & - \\
    list/sum/2\_original & - & - & - \\
    list/sum/3\_assisted & - & - & - \\
    list/sum/3\_original & - & - & - \\
    nat/assoc/left1 & - & - & \checkmark \\
    nat/assoc/left2 & - & - & \checkmark \\
    nat/assoc/right1 & - & - & \checkmark \\
    nat/assoc/right2 & - & - & - \\
    nat/bound\_of\_sum/lower\_bound\_of\_sum\_assisted & \checkmark & - & \checkmark \\
    nat/bound\_of\_sum/lower\_bound\_of\_sum\_original & - & - & \checkmark \\
    nat/bound\_of\_sum/strict\_lower\_bound\_of\_sum\_assisted & - & - & \checkmark \\
    nat/bound\_of\_sum/strict\_lower\_bound\_of\_sum\_original & - & - & \checkmark \\
    nat/bound\_of\_sum/strict\_upper\_bound\_of\_sum\_assisted & - & - & - \\
    nat/bound\_of\_sum/strict\_upper\_bound\_of\_sum\_original & - & - & - \\
    nat/bound\_of\_sum/upper\_bound\_of\_sum\_assisted & - & - & - \\
    nat/bound\_of\_sum/upper\_bound\_of\_sum\_original & - & - & - \\
    nat/dist/left1\_assisted & - & - & - \\
    nat/dist/left1\_original & - & - & - \\
    nat/dist/left2\_assisted & - & - & - \\
    nat/dist/left2\_original & - & - & - \\
    nat/dist/left3\_assisted & - & - & - \\
    nat/dist/left3\_original & - & - & - \\
    nat/dist/right1\_assisted & - & - & - \\
    nat/dist/right1\_original & - & - & - \\
    nat/dist/right2\_assisted & - & - & - \\
    nat/dist/right2\_original & - & - & - \\
    nat/dist/right3\_assisted & - & - & - \\
    nat/dist/right3\_original & - & - & - \\
    nat/division/assisted & - & - & - \\
    nat/division/original & - & - & - \\
    nat/divisor/1\_assisted & - & - & - \\
    nat/divisor/1\_original & - & - & - \\
    nat/divisor/2\_assisted & - & - & - \\
    nat/divisor/2\_original & - & - & - \\
    nat/double/1 & - & - & \checkmark \\
    nat/double/2 & - & - & \checkmark \\
    nat/double/3 & - & - & - \\
    nat/double/4 & - & - & - \\
    nat/factorial/1 & - & - & - \\
    nat/factorial/2 & - & - & - \\
    nat/fibonacci/1 & - & - & - \\
    nat/fibonacci/2 & - & - & - \\
    nat/gcd/1\_assisted & - & - & - \\
    nat/gcd/1\_original & - & - & - \\
    nat/gcd/2\_assisted & - & - & - \\
    nat/gcd/2\_original & - & - & - \\
    nat/gcd/3\_assisted & - & - & - \\
    nat/gcd/3\_original & - & - & - \\
    nat/predecessor & \checkmark & - & \checkmark \\
    nat/quo-rem/quo-rem\_original & - & - & - \\
    nat/quo-rem/quo\_assisted & - & - & - \\
    nat/quo-rem/quo\_original & - & - & - \\
    nat/quo-rem/rem\_assisted & - & - & - \\
    nat/quo-rem/rem\_original & - & - & - \\
    nat/root/assisted & - & - & - \\
    nat/root/original & - & - & - \\
    nat/subtraction/leq\_left\_assisted & - & - & - \\
    nat/subtraction/leq\_left\_original & - & - & - \\
    nat/subtraction/leq\_right\_assisted & - & - & - \\
    nat/subtraction/leq\_right\_original & - & - & - \\
    nat/subtraction/less\_left\_assisted & - & - & - \\
    nat/subtraction/less\_left\_original & - & - & - \\
    nat/subtraction/less\_right\_assisted & - & - & - \\
    nat/subtraction/less\_right\_original & - & - & - \\
    \hline
\end{longtable}
\end{center}

%% file: paper.bbl
\begin{thebibliography}{10}
\providecommand{\url}[1]{\texttt{#1}}
\providecommand{\urlprefix}{URL }
\providecommand{\doi}[1]{https://doi.org/#1}

\bibitem{sygus-comp}
Alur, R., Fisman, D., Padhi, S., Reynolds, A., Singh, R., Udupa, A.:
  {SyGuS-Comp 2019} (2019), \url{https://sygus.org/comp/2019/}

\bibitem{cvc5}
Barbosa, H., Barrett, C.W., Brain, M., Kremer, G., Lachnitt, H., Mann, M.,
  Mohamed, A., Mohamed, M., Niemetz, A., N{\"{o}}tzli, A., Ozdemir, A.,
  Preiner, M., Reynolds, A., Sheng, Y., Tinelli, C., Zohar, Y.: {cvc5: {A}
  Versatile and Industrial-Strength {SMT} Solver}. In: TACAS. pp. 415--442
  (2022). \doi{10.1007/978-3-030-99524-9\_24}

\bibitem{SMTLIB2.7}
Barrett, C., Fontaine, P., Tinelli, C.: {The SMT-LIB Standard: Version 2.7}.
  Tech. rep., Department of Computer Science, The University of Iowa (2025),
  \url{https://smt-lib.org/papers/smt-lib-reference-v2.7-r2025-02-05.pdf}

\bibitem{Z3}
De~Moura, L., Bj{\o}rner, N.: {Z3: An Efficient SMT Solver}. In: TACAS. LNCS,
  vol.~4963, pp. 337--340. Springer (2008). \doi{10.1007/978-3-540-78800-3\_24}

\bibitem{LemmaDiscoveryAndStrategies}
Einarsdóttir, S.H., Hajdu, M., Johansson, M., Smallbone, N., Suda, M.: {Lemma
  Discovery and Strategies for Automated Induction}. In: IJCAR (2024).
  \doi{10.1007/978-3-031-63498-7_13}

\bibitem{ProgramSynthesis}
Gulwani, S., Polozov, O., Singh, R.: Program synthesis. Foundations and
  Trends® in Programming Languages  \textbf{4}(1-2),  1--119 (2017).
  \doi{10.1561/2500000010}

\bibitem{RecSynSat}
Hozzov{\'a}, P., Amrollahi, D., Hajdu, M., Kov{\'a}cs, L., Voronkov, A.,
  Wagner, E.M.: {Synthesis of Recursive Programs in Saturation}. In:
  Benzm{\"u}ller, C., Heule, M.J., Schmidt, R.A. (eds.) Automated Reasoning.
  pp. 154--171. Springer Nature Switzerland, Cham (2024).
  \doi{10.1007/978-3-031-63498-7_10}

\bibitem{SynSat}
Hozzov{\'a}, P., Kov{\'a}cs, L., Norman, C., Voronkov, A.: {Program Synthesis
  in Saturation}. In: CADE. pp. 307--324 (2023).
  \doi{10.1007/978-3-031-38499-8_18}

\bibitem{synthesiz3}
Hozzová, P., Bjørner, N.: {Synthesiz3 This: an SMT-Based Approach for
  Synthesis with Uncomputable Symbols} (2025),
  \url{https://arxiv.org/abs/2504.16536}, to appear in proceedings of FMCAD
  2025.

\bibitem{CAV13}
Kov{\'a}cs, L., Voronkov, A.: {First-Order Theorem Proving and Vampire}. In:
  CAV. pp. 1--35 (2013)

\bibitem{MannaWaldinger1980}
Manna, Z., Waldinger, R.: {A Deductive Approach to Program Synthesis}. ACM
  Trans. Program. Lang. Syst.  \textbf{2}(1),  90–121 (1980).
  \doi{10.1145/357084.357090}

\bibitem{MannaWaldinger1971}
Manna, Z., Waldinger, R.J.: {Toward automatic program synthesis}. Commun. ACM
  \textbf{14}(3),  151–165 (Mar 1971). \doi{10.1145/362566.362568}

\bibitem{sygus-standard}
Padhi, S., Polgreen, E., Raghothaman, M., Reynolds, A., Udupa, A.: {The SyGuS
  Language Standard Version 2.1} (2021),
  \url{https://sygus-org.github.io/language/}

\bibitem{Reynolds19}
Reynolds, A., Kuncak, V., Tinelli, C., Barrett, C.W., Deters, M.:
  {Refutation-based synthesis in {SMT}}. Formal Methods Syst. Des.
  \textbf{55}(2),  73--102 (2019). \doi{10.1007/S10703-017-0270-2}

\end{thebibliography}
